\documentclass[namedreferences,psfig]{kluwer}

\usepackage{epsfig}

\def\beq{\begin{equation}}
\def\eeq{\end{equation}}
\def\bea{\begin{eqnarray}}
\def\eea{\end{eqnarray}}

\def\msol{{M}_{\odot}}
\def\ltsima{$\; \buildrel < \over \sim \;$}
\def\simlt{\lower.5ex\hbox{\ltsima}}

\begin{document}
\begin{article}
\begin{opening}

\title{$\Omega_m$ -- Different Ways to Determine
the Matter Density of the Universe}

\author{Sabine \surname{Schindler}\email{sas@astro.livjm.ac.uk}}

\institute{Astrophysics Research Institute, Liverpool John Moores University,\\
Twelve Quays House, Birkenhead CH41 1LD, U.K.}

\begin{abstract}

A summary of various measurements of the mean matter density in the 
universe, $\Omega_m$, is presented. Results from very different kinds of
methods using various
astronomical objects -- from supernovae to large-scale structure --
are shown. There is  a remarkable preference for  $\Omega_m$ values
around
0.3, but there are also some measurements that favour a higher or 
a smaller value.

\end{abstract}
\end{opening}

\section{Introduction}

$\Omega_m$ -- the mean matter density in the universe 
-- is one of
the key parameters for cosmological models. 
It is usually expressed as a fraction of the critical density

$$\Omega_m = {\rho_m \over \rho_{crit}}$$
with $\rho_{crit} = 3H_0^2 / (8\pi G) = 1.88h^2 \times 10^{-29}
g/cm^3$.
Several years ago the
philosophically appealing value of $\Omega_m=1$ was favoured by most
scientists. But this value led more and more to contradictions with
various other measurements, which require a lower value.

Recently, many
different methods using different kinds of astronomical objects 
have been developed to determine $\Omega_m$.
In this article measurements of the matter density with very different 
methods and their results are summarised. 
It is impossible to present a complete compilation of all results on
this topic due to the limited space. Therefore only a selection of
methods and results are presented, whereby I concentrate on
recent determinations of $\Omega_m$. 

This article is organised as
follows. In Sect.~2 the combined results from distant supernovae and
measurements of the cosmic microwave background radiation are
given. Sect.~3 lists $\Omega_m$ values determined with the gravitational
lensing effect. In Sects.~4 and 5 the evolution and the baryon fraction of
galaxy clusters, respectively, are used to determine the matter
density. The results from mass-to-light ratios are listed in
Sect.~6. $\Omega_m$ determinations from cosmic flows, and from  correlation
functions and power spectra 
are given in Sect.~7 and 8, respectively. Sect.~9 summarises
all the results obtained by the different methods.   
Throughout this article a
Hubble constant of $H_0=65$ km/s/Mpc is used.

\section{Supernovae and Cosmic Microwave Background}

Currently, the most discussed results for $\Omega_m$ are derived
from a combination of supernova and Cosmic Microwave Background
measurements. 

Recent measurements of  the Cosmic
Microwave Background Radiation (CMBR) determined the small-angle 
anisotropies of this
radiation over a significant part of the sky.
The angular power spectrum of these measurements yields 
values for the total density $\Omega_{tot}$
around unity. Two balloon experiments  find the following
results: Boomerang (de Bernardis et al. 2000)

$$0.88<\Omega_{tot}<1.12$$
and MAXIMA (Hanany et al. 2000, Balbi et al. 2000)

$$\Omega_{tot}=1.0^{+0.15}_{-0.30}.$$

Distant Type Ia supernovae can be used as standard candles and hence
they can be used to determine cosmological parameters.
With the assumption of the total density
$\Omega_{tot}=\Omega_m+\Omega_{\Lambda}=1$ -- as suggested by the CMBR
measurements -- quite stringent 
constraints can be set on the matter density. Two independent
groups measured supernovae for this purpose and found very similar results:

$$\Omega_m = 
0.28^{+0.09}_{-0.08}\hskip 0.2cm{\rm Perlmutter~ et~ al.~ (1999)}$$
and

$$\Omega_m = 0.32\pm0.1\hskip 0.2cm {\rm Riess~ et~ al.~ (1998)}.$$
The main concerns about the interpretation of these data are the
evolution of supernovae Ia and dimming by dust.

 
\section{Gravitational lensing}

There are several ways to determine $\Omega_m$ by the
gravitational lensing effect. A very interesting method is the
weak gravitational lensing by of large-scale structure -- the cosmic shear.
Four independent groups discovered the effect recently (Bacon et al. 2000;
Kaiser et al. 2000; van Waerbeke et al. 2000; Wittman et al. 2000).
The first values for $\Omega_m$ were given by  van Waerbeke et al. (2001)

$$0.2 < \Omega_m < 0.5 {\rm ~~~~for~ an~ open~ universe}$$
and

$$\Omega_m < 0.4 {\rm  ~~~~for~ a~ flat~ universe}.$$
Wilson et al. (2001) determined mass-to-light ratios from the 
gravitational shear of many faint galaxies implying a very low value for 

$$\Omega_m = 0.10\pm0.02.$$

Another method is arc statistics, i.e. the
number of giant gravitational
arcs caused by lensing of foreground objects.
X-ray selected clusters are ideal objects for this purpose. 
Bartelmann et al. (1998)  and Kaufmann \& Straumann (2000)
applied this method to the cluster sample of the EINSTEIN
Medium Sensitivity Survey (EMSS) (Gioia \& Luppino 1994; Luppino et
al. 1999).
Kaufmann \& Straumann (2000) derive with semi-analytic methods 
a mean matter density
between 

$$0.2 < \Omega_m < 0.5.$$
Bartelmann et al. (1998) find from numerical simulations also a low
value for $\Omega_m$.
In principle also the arcs found in  radio surveys can determine $\Omega_m$
(Helbig 2000). 
The CLASS/JVAS surveys already give some results, but up to now
only relatively weak constraints can be set on $\Omega_m$.

A new method was suggested by Golse et al. (2001). They show that
strong lensing in galaxy clusters with several
image systems can constrain cosmological parameters without any
further assumptions. When high resolution images and the redshifts of
the gravitational arcs are available a single galaxy
cluster with 3 multiple image
systems can determine $\Omega_m$ with an uncertainty of about $\pm 0.3$.

\section{Cluster evolution}

For $\Omega_m = 1$ a strong evolution is
predicted in the number space density of rich clusters, because in
this cosmological model the growth of
structure continues to the present day. In a low-$\Omega$ universe, on the
other hand, relatively
little change in the cluster number is expected since a redshift of  1. 
Therefore much work has been done to test whether there is evolution
or not in the cluster number density.

\subsection{Single Clusters}

The existence of a single distant, massive cluster -- 
MS1054-03 at a redshift of z=0.83 with a mass $M \approx
10^{15}\msol$ -- is by itself 
a strong indication for a low $\Omega_m$ universe
(Donahue et al. 1998). This cluster and two more clusters were used in
an analysis 
by Bahcall \& Fan (1998), in which they also found a low value

$$\Omega_m=0.2^{+0.3}_{-0.1}.$$

\subsection{X-ray Luminosity Function}

To put this type of analysis on a broader statistical basis 
the evolution of the cluster mass function, i.e. the evolution of
the number of clusters of different masses, would be the ideal quantity 
to measure. But as it is not easy to determine the mass for a  large
number of clusters, 
the cluster luminosity function and the cluster temperature function 
have generally been studied instead. This is possible because both 
quantities -- the X-ray luminosity and the temperature --
correlate quite well with the cluster mass.

\begin{table}
\caption{Various projects evaluating the cluster X-ray luminosity
function which find a deficit of high-redshift clusters. The acronym  
of the project (Column 2) and the significance of deficit of distant luminous
clusters (Column 3) are listed . The study by Gioia \& Luppino (1994) 
is based on the EINSTEIN Medium Sensitivity Survey. 
All the others are based on ROSAT data.}
\label{tab1}
\begin{tabular}{|l|c|c|} \hline
Gioia \& Luppino (1994)  &                & $3\sigma$ \cr
Nichol et al. (1999)            & SHARC &$1.7\sigma$ \cr
Vikhlinin et al. (2000)        &                &$3.5\sigma$ \cr
Rosati et al. (2000)             & RDCS    &$3\sigma$ \cr
Gioia et al. (2001)               &  NEP      &$5\sigma$\cr
\hline
\end{tabular}
\end{table}

The luminosity function of 
X-ray selected clusters has been measured by several groups. In many 
measurements a deficit of distant luminous clusters (see Table~1)
was found.
These results point towards a high $\Omega_m$ universe, but the current
results have still a large uncertainty and therefore they 
cannot exclude an $\Omega_m=0.3$.
There is one analysis that yields a different result although it is based 
on the same data from ROSAT: Jones et al. (2000) found no deficit of
distant clusters. Hence 
they concluded that there is ``no evolution'' in the cluster luminosity
function, which is an indication for a low $\Omega_m$ universe.

\subsection{Temperature Function}

Several groups have investigated the temperature function of galaxy 
clusters and found discordant results. Evidence for ``no evolution''
was found by  Eke et al. (1998). They determined a matter density of

$$\Omega_m = 0.45 \pm 0.25.$$
Henry (2000) also did not find any evolution and concluded  

$$\Omega_m = 0.49\pm0.12 {\rm ~~~~for ~an ~open ~universe}$$

$$\Omega_m = 0.44\pm0.12  {\rm ~~~~for ~a ~flat ~universe}.$$

Two other groups found evidence that there is evolution in the
temperature function. 
Viana \& Liddle (1999) derived

$$\Omega_m = 0.75 \pm 0.3$$
and Blanchard et al. (2000) found 

$$\Omega_m = 0.92^{+0.26}_{-0.22} {\rm ~~~~for ~an ~open ~universe}$$

$$\Omega_m = 0.87^{+0.35}_{-0.25} {\rm ~~~~for ~a ~flat ~universe}$$

Maybe the sample selection must be done more carefully in 
order to find agreement. It also might be that 
the temperature function is only a weak
test in the redshift range used here as it was suggested by  
Colafrancesco et al. (1997).

\subsection{Mass Function}

The evolution of the mass function has been measured directly by 
Carlberg et al. (1997a) with the CNOC (Canadian Network for
Observational Cosmology) sample.
The clusters in this sample were selected from the EMSS. 
The masses were obtained from optical measurements
of the galaxy velocity dispersion. Carlberg et al. (1997a)
find a low value for $\Omega_m$:

$$\Omega_m=0.2\pm0.1.$$

\subsection{X-ray luminosity -- Temperature Relation}

The evolution of the X-ray luminosity -- temperature relation is
another test for the mean matter density, because it evolves
differently in different cosmological models. Several authors
concluded 
that there is no significant detectable evolution in the relation: 
Mushotzky \& Scharf (1997) for a sample out to redshift $z\simlt 0.4$,
Donahue et al. (1998) and Della Ceca et al. (2000) out to $z\simlt 0.8$,
Schindler (1999) out to $z\simlt 1.0$,
Fabian et al. (2001) out to $z\simlt 1.8$. From a detailed comparison
of the ROSAT Deep Cluster Survey (Rosati et al. 1995) and the
EMSS Sample Borgani et al. (1999) derived 

$$\Omega_m = 0.4^{+0.3}_{-0.2} {\rm ~~~~for ~an ~open ~universe}$$
and

$$\Omega_m \simlt 0.6 {\rm ~~~~for ~a ~flat ~universe.}$$

\section{Cluster baryon fraction}

With the assumption that the matter is accumulated 
indiscriminately in the potential
wells of clusters the baryon fraction in galaxy clusters
is a measure for the baryon fraction of the universe as a whole. The
advantage of measuring 
the baryon fraction in clusters is that both the baryon mass and the
total  cluster mass can be determined reliably (Schindler 1996). 
For the analysis only
the gas density and the gas temperature are required which can both
be inferred from X-ray observations. Several groups determined
gas mass fractions from X-ray observations in samples of nearby and 
distant clusters, e.g. 

\begin{itemize}
\item Mohr et al. (1999): \hskip 1.6cm $f_{gas}=0.14$ 
\item Ettori \& Fabian (1999):   \hskip .8cm $f_{gas}=0.11$
\item Arnaud \& Evrard (1999): \hskip .5cm $f_{gas}=0.12$
\item Schindler (1999):  \hskip 1.9cm $f_{gas}=0.12$
\end{itemize}
All these determinations depend on the radius where the mass fraction
is determined, because the gas mass fraction increases slightly 
with radius. In the above mentioned analyses the mass was determined
within a radius $r_{500}$ from the cluster centre. 
This radius encompasses a volume 
that has a density of 500 $\times$ the
critical density of the universe $\rho_{crit}$. 
Out to this radius the X-ray profile
necessary for the analysis could be measured reliably.

To determine $\Omega_m$, the gas mass fraction $f_{gas}$ must
be compared to the baryon density in the universe
$\Omega_B \simlt 0.05$ determined  
from primordial nucleosynthesis (see e.g. Burles \& Tytler
1998a,b). The ratio of the baryon density and the gas mass fraction
yields an upper limit for the matter density $\Omega_m$:

$$\Omega_m < {\Omega_B\over f_{gas}} = 0.3 - 0.4$$

The baryon fraction can also be determined in a different way: 
measurements of  the Sunyaev-Zel'dovich effect -- inverse-Compton
scattering of the Cosmic Microwave Background photons by the
hot intra-cluster gas shifts the CMBR spectrum to slightly higher energies. 
As this effect is proportional to the gas density,
the density profile can be determined directly. 
Only an additional measurement
of the gas temperature is necessary from X-rays. 
The gas mass fraction found

\begin{itemize}
\item Grego et al. (2001): \hskip .5cm $f_{gas}=0.13$ 
\end{itemize}
is very similar to the X-ray results.
Hence they derive also a similar upper limit for the matter density

$$ \Omega_m < 0.4.$$
In these analyses only the mass in the intra-cluster gas was taken into
account. Baryons in the galaxies were neglected. If they were to be
included, 
the baryon fraction would increase  slightly and hence ever more stringent
constraints on $\Omega_m$ could be placed.

\section{Mass-to-light ratio}

The matter density in the universe $\Omega_m$ is defined as the ratio
of the 
mean matter density $\rho_m$ and the critical density $\rho_{crit}$ 

$$\Omega_m = {\rho_m \over \rho_c} = {M \over L} {j \over \rho_{crit}}.$$
$\rho_m$ can also be expressed as the mass - to -(optical)light ratio times
the field luminosity density $j$. The 
assumption here is that  mass-to-light ratios in galaxy clusters are 
representative for the whole universe. This is probably a good assumption 
because clusters draw mass and galaxy content from regions of about 40 Mpc
in size.

Carlberg et al. (1997b) inferred mass-to-light ratios from the CNOC sample.
They could also measure directly with
their data the value for the field luminosity density.
The resulting matter density is

$$\Omega_m = 0.19\pm0.06.$$

From a  comparison of cosmological hydrodynamic simulations by Cen \&
Ostriker (1999), and observations,
Bahcall et al. (2000) also determined mass-to-light ratios and
concluded that the matter density is

$$\Omega_m = 0.16\pm0.05.$$

\section{Cosmic flows}

Measurements of  peculiar velocities of galaxies and clusters on
large scales can be used to determine the large-scale potential and
hence the mass content of the universe. In linear perturbation
theory there is a linear relation between the peculiar velocity
and the gravity field. The only uncertainty is the 
proportionality factor 
$\beta = {\Omega_m^{0.6} \over b}$ -- the biasing, which reflects
that the visible matter does not exactly trace the total matter.
Unfortunately, 
is factor is up to now not very well defined (see e.g. Strauss 1999).

Many groups determined $\Omega_m$ from these cosmic flows. Some of the results 
are summarised in Table~2 (see also Fig.~1). Although for the various analyses 
the same catalogues were used very different results were
obtained. The reason for the discrepancies is probably
the uncertainty in the biasing parameter.

\begin{table}
\caption{$\Omega_m$ - values derived by several groups. Column (2)
lists the catalogues used: MARK III and SFI are catalogues of
galaxies, Abell is a catalogue of galaxy clusters.}
\label{tab2}
\begin{tabular}{|l|c|c|} \hline
Willick \& Strauss (1998) & MARK III& $\Omega_m \approx 0.3$ \cr
Susperregi (2001)               & MARK III&$\Omega_m \approx 0.3$ \cr
Zaroubi et al. (1997)           & MARK III&$\Omega_m = 0.5\pm0.1$ \cr
Freudling et al. (1999)        & SFI            &$\Omega_m = 0.5\pm0.1$ \cr
Bridle et al. (2001)               & SFI            &$0.25<\Omega_m <0.89$ \cr
Sigad et al. (1998)               & MARK III&$\Omega_m \approx 1$ \cr
Branchini et al. (2000)        & Abell         &$\Omega_m \approx 1$ \cr
\hline
\end{tabular}
\end{table}

\section{Correlation functions and power spectra}

As was shown by Mo et al. (1996) the cluster correlation function 
can be used to determine $\Omega_m$. Different cluster
samples have been used: optically selected clusters 
(Croft et al. 1997: APM)
and X-ray selected samples (Moscardini et al. 2000; Collins
et al. 2000; Schuecker et al. 2001: ROSAT). All analyses
favour a low $\Omega_m$, but no ranges for $\Omega_m$ are given so far
because the constraints are not very stringent yet.

The power spectrum of the Ly$\alpha$
forest was used by Croft et al. (1999). The authors find a 
matter density of 

$$\Omega_m \approx 0.4.$$
Weinberg et al. (1999) combined galaxy clusters and measurements of
the  Ly$\alpha$
forest. They adopted a shape parameter of the power spectrum 
$\Gamma=0.2$ which is favoured by a number of
studies of large-scale galaxy clustering. Their results for the matter
density are 

$$\Omega_m = 0.46^{+0.12}_{-0.10} {\rm ~~~~for ~an  ~open ~universe}$$
and

$$\Omega_m = 0.34^{+0.13}_{-0.09} {\rm ~~~~for ~a ~flat ~universe}.$$

\section{Summary on $\Omega_m$ }

\begin{figure}
\epsfig{figure=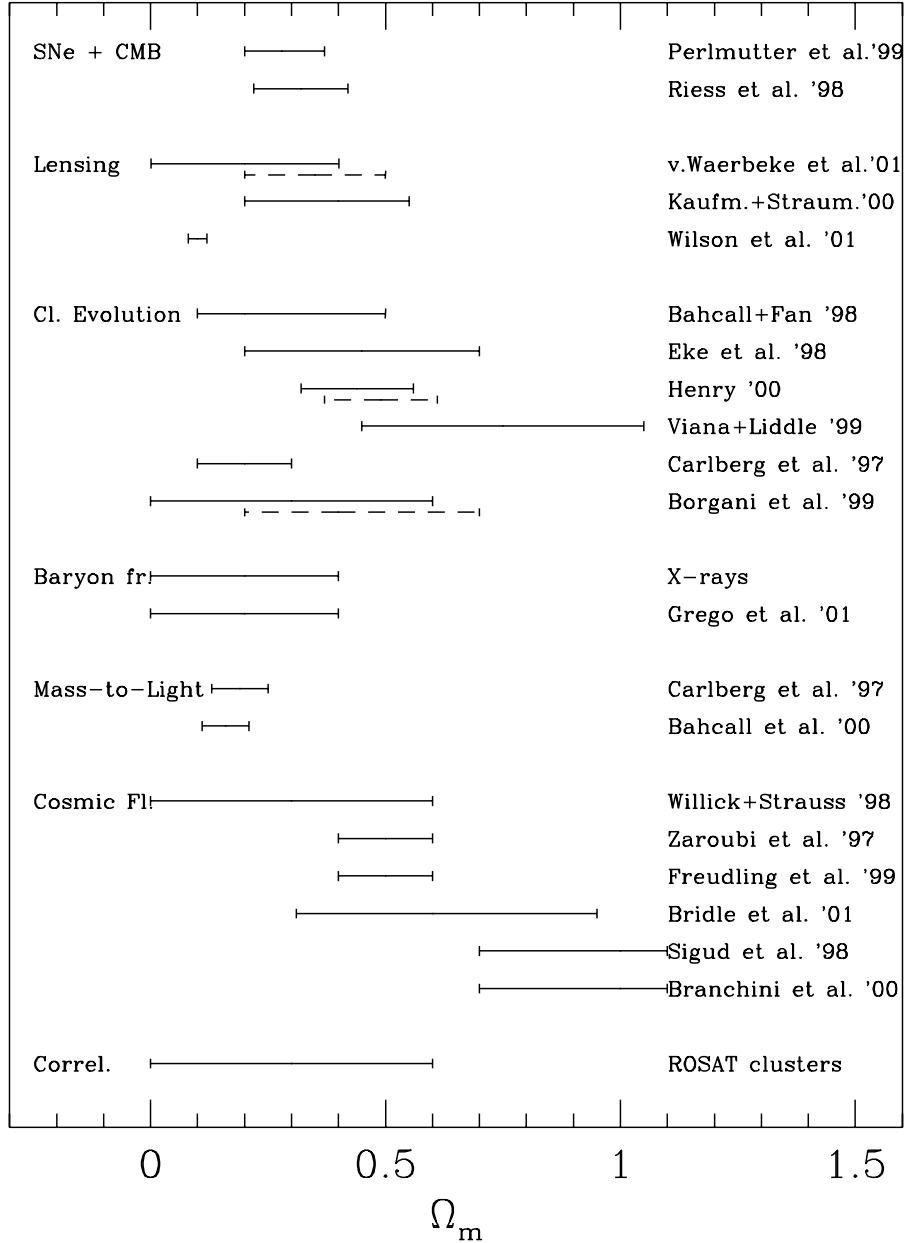,width=12cm,clip=}
\caption[]{Summary of $\Omega_m$ values as derived by the different
methods (selection only). 
The methods are listed on the left hand side, the references
on the right hand side. If the authors distinguished between open and
flat models, the flat models are shown as full lines and the open
models as dashed lines. Most methods are in agreement with an
$\Omega_m \approx 0.3$, but there are also some results that favour
other values.
}
\end{figure}

An overview of results for $\Omega_m$ obtained with the different
methods is shown in Fig.~1. Unfortunately, it is
impossible to plot all the results in one diagram 
because of  the large number of  publications on this topic.
Therefore only a selection of results is shown.
The diagram is simplified in the sense that for some methods assumptions
had to be made which cannot be shown in such a simple figure. Some authors
do not state error ranges, therefore for some data points errors
had to be assumed. 

Some publications distinguish between open and flat models. In these
cases the flat models are shown as full lines and the open models as 
dashed lines. There is a systematic shift between the results for these two
models. Therefore the region constrained e.g. by galaxy clusters is not a
vertical bar in $\Omega_m - \Omega_{\Lambda}$ diagram but a bar
slightly tilted
towards the line defining $\Omega_m + \Omega_{\Lambda} = 1$.

As shown in Fig.~1 most values cluster around $\Omega_m=0.3$. 
There is a 
remarkable agreement between determinations from completely different
methods using various astronomical objects from supernovae to the mass
distribution on large scales. But there are also
measurements that 
favour higher or lower values. It is not that a particular method
yields systematically lower or higher values, but it seems rather a
scatter which depends on certain assumptions made by the authors or
simply on large uncertainties in the measurements. Therefore new and future
observational facilities, e.g. CHANDRA, XMM and PLANCK,  will ensure 
that the coming years remain exciting for cosmology.
           
\begin{acknowledgements}
I would like to thank Phil James for carefully reading the manuscript.
\end{acknowledgements}

{}

\end{article}

\begin{thebibliography}{}

\bibitem{} Arnaud M., Evrard A.E., 1999, MNRAS 305, 631

\bibitem{} Bacon D.J., Refregier A.R., Ellis R.S., 2000, MNRAS 318, 625
 
\bibitem{} Bahcall N.A., Fan X., 1998, ApJ 504, 1

\bibitem{} Bahcall N.A., Cen R., Dav\'e R., Ostriker J.P., Yu Q.,
2000, ApJ 541, 1

\bibitem{} Balbi A., Ade P., Bock J., et al., 2000, ApJ 545, L1

\bibitem{} Bartelmann M., Huss A., Colberg J.M., 1998, A\&A 330, 1

\bibitem{} Blanchard A., Sadat R., Bartlett J.G., Le Dour M., 2000,
A\&A 362, 809

\bibitem{} Borgani S., Rosati P., Tozzi P., Norman C., 1999, ApJ 517, 40

\bibitem{} Branchini E., Zehavi I., Plionis M., Dekel A., 2000, MNRAS
313, 491

\bibitem{} Bridle S.L., Zehavi I., Dekel A., et al., 2001, MNRAS 321, 333

\bibitem{} Burles S., Tytler D., 1998a, ApJ 499, 699

\bibitem{} Burles S., Tytler D., 1998b, ApJ 507, 732

\bibitem{} Carlberg R.G., Morris S.L., Yee H.K.C., Ellingson E., 1997a,
ApJ 479, L19 

\bibitem{} Carlberg R.G., Yee H.K.C., Ellingson E., 1997b, ApJ 478, 462

\bibitem{} Cen R., Ostriker J.P., 1999, ApJ 514, 1

\bibitem{} Colafrancesco S., Mazzotta P., Vittori N., 1997, ApJ 488, 566

\bibitem{} Collins C.A., Guzzo L., B\"ohringer H., et al., 
2000, MNRAS 319, 939

\bibitem{} Croft R.A.C., Dalton G.B., Efstathiou G., Sutherland W.J.,
Maddox S.J., 1997, MNRAS 291, 305

\bibitem{} Croft R.A.C., Weinberg D.H., Pettini M., Hernquist L., Katz
N., 1999, ApJ 520, 1

\bibitem{} De Bernardis P., Ade P.A.R., Bock J.J., et al., 2000, Nature
404, 955

Della Ceca R., Scaramella R., Gioia I.M., et al., 2000, A\&A 353, 498

\bibitem{} Donahue M., Voit G.M., Gioia I., et al., 1998, ApJ 502, 550

\bibitem{} Eke V.R., Cole S., Frenk C.S., Henry J.P., 1998, MNRAS 298, 1145 

\bibitem{} Ettori S., Fabian A.C., 1999, MNRAS 305, 834

\bibitem{} Fabian A.C., Crawford C.S., Ettori S.,
 Sanders J.S., 2001, MNRAS 322, L11

\bibitem{} Freudling W., Zehavi I., Da Costa L.N., 1999, ApJ 523, 1

\bibitem{} Gioia I.M., Luppino G.A., 1994, ApJS 94, 583

\bibitem{} Gioia I.M., Henry, J.P., Mullis C.R., et al., 2001, ApJ
553, L105

\bibitem{} Golse G., Kneib J.-P., Soucail G., 2001, astro-ph/0103500

\bibitem{} Grego L., Carlstrom J.E., Reese E.D., et al., 2001, ApJ
552, 2

\bibitem{} Hanany S., Ade P., Balbi A., et al., 2000, ApJ 545, L5

\bibitem{} Helbig P., 2000, Proceedings of IAU Symposium 201 `New
Cosmological Data and the Values of the Fundamental Parameters',
A. Lasenby and A. Wilkinson (eds.), astro-ph/0011031

\bibitem{} Henry J.P., 2000, ApJ 534, 565

\bibitem{} Jones L.R., Ebeling H., Scharf C., etal.,  2000,
``Large Scale Structure in the X-ray Universe'', Plionis, M., 
Georgantopoulos, I. (eds.), Atlantisciences, Paris, France, p.35

\bibitem{} Kaiser N., Wilson G., Luppino G., 2000, astro-ph/0003338

\bibitem{} Kaufmann R., Straumann N., 2000, Ann.Phys. 9, 384

\bibitem{} Luppino G.A., Gioia I.M., Hammer F., et al., 1999, A\&AS,
136, 117


\bibitem{} Mo H.J., Jing Y.P., White S.D.M., 1996, MNRAS 282, 1096

\bibitem{} Mohr J.J., Mathiesen B., Evrard A.E., 1999, ApJ 517, 627

\bibitem{} Moscardini L., Matarrese S., De Grandi S., Lucchin F.,
2000, MNRAS 314, 647

\bibitem{} Mushotzky R.F., Scharf C.A., 1997, ApJ 482, L13 


\bibitem{} Nichol R.C., Romer A.K., Holden B.P., et al., 1999, ApJ 521,
L21

\bibitem{} Perlmutter S., Aldering G., Goldhaber G., et al., 1999, ApJ
517, 565

\bibitem{} Riess A.G., Filippenko A.V., Challis P., et al. 1998, ApJ
116, 1009

\bibitem{} Rosati P., Della Ceca R., Burg R., Norman C., Giacconi R.,
1995, ApJ 445, L11

\bibitem{} Rosati P., Borgani S., Della Cecca R., et al., 2000,
``Large Scale Structure in the X-ray Universe'', Plionis, M., 
Georgantopoulos, I. (eds.), Atlantisciences, Paris, France, p.13

\bibitem{} Schindler S., 1996, A\&A 305, 756

\bibitem{} Schindler S., 1999, A\&A 349, 435

\bibitem{} Schuecker P., B\"ohringer H., Guzzo L., et al., 2001, A\&A
368, 86 

\bibitem{} Sigad Y., Eldar A., Dekel A., Strauss M.A., Yahil A., 1998,
ApJ 495, 516

\bibitem{} Strauss M., 1999, ``Cosmic Flows Workshop'', S.
Courteau, M. Strauss \& J. Willick (eds.), ASP series, astro-ph/9908325

\bibitem{} Susperregi M., 2001, ApJ 546, 85

\bibitem{} Van Waerbeke L., Mellier Y., Erben T., et al., 2000, A\&A 358, 30

\bibitem{} Van Waerbeke L., Mellier Y., Radovich M., et al., 
2001, astro-ph/0101511

\bibitem{} Viana P.T.P., Liddle A.R., 1999, MNRAS 303, 535

\bibitem{} Vikhlinin A., McNamara B., Quintana H., et al., 2000, 
``Large Scale Structure in the X-ray Universe'', Plionis, M., 
Georgantopoulos, I. (eds.), Atlantisciences, Paris, France, p.31

\bibitem{} Weinberg D.H., Croft R.A.C., Hernquist L., Katz N., Pettini
M., 1999, ApJ 522, 563

\bibitem{} Willick J.A., Strauss M.A., 1998, ApJ 507, 64

\bibitem{} Wilson G., Kaiser N., Luppino G.A., 2001, astro-ph/0102396

\bibitem{} Wittman D.M., Tyson J.A., Kirkman D., et al., 2000, Nature
405, 143

\bibitem{} Zaroubi S., Zehavi I., Dekel A., Hoffman Y., Kolatt T.,
1997, ApJ 486, 21

\end{thebibliography}
\end{document}